\documentclass[aps,prd,superscriptaddress,notitlepage,showkeys,showpacs,10pt]{revtex4-1}

\usepackage{graphicx}
\usepackage{amsmath}
\usepackage{amssymb}
\usepackage{color}
\usepackage{bm}
\usepackage{hyperref}
\usepackage{float}

\definecolor{purple}{rgb}{0.8,0,0.6}

\begin{document}
\title{Magnetogenesis during inflation and preheating in the Starobinsky model}

\author{S.~Vilchinskii}
\affiliation{Department of Physics, Taras Shevchenko National University of Kyiv, Kyiv, 03022, Ukraine}
\affiliation{D\'{e}partement de Physique Th\'{e}orique, Center for Astroparticle Physics, Universit\'{e} de Gen\`{e}ve, 1211 Gen\`{e}ve 4, Switzerland}

\author{O.~Sobol}
\affiliation{Department of Physics, Taras Shevchenko National University of Kyiv, Kyiv, 03022, Ukraine}

\author{E.~V.~Gorbar}
\affiliation{Department of Physics, Taras Shevchenko National University of Kyiv, Kyiv, 03022, Ukraine}
\affiliation{Bogolyubov Institute for Theoretical Physics, Kyiv, 03680, Ukraine}

\author{I.~Rudenok}
\affiliation{Department of Physics, Taras Shevchenko National University of Kyiv, Kyiv, 03022, Ukraine}

\begin{abstract}
By assuming the kinetic coupling $f^{2}(\phi)FF$ of the effective inflaton field $\phi$ with the electromagnetic field, we explore 
magnetogenesis during the inflation and preheating stages in the $R^2$ Starobinsky model \cite{Starobinsky}.  We consider the case of the exponential coupling function
$f(\phi)=\exp(\alpha\phi/M_{p})$ and show that for $\alpha \sim 12-15$ it is possible to generate the large scale magnetic fields with 
strength $\gtrsim 10^{-15}$~Gauss at the present epoch. 
The spectrum 
of generated magnetic fields is blue with the spectral index $n=1+s$, $s>0$. 
We have found 
that for the relevant values of the coupling parameter, $\alpha=12-15$,  model avoids the back-reaction problem for all relevant modes. 
\end{abstract}

\maketitle

\section{Introduction}

A variety of observations imply that stars, galaxies, and clusters of galaxies are all magnetized. The typical magnetic field strengths range 
from few $\mu{\rm G}$ in the case of galaxies and galaxy clusters up to $10^{15}$ G in magnetars (see, e.g, Refs.~
\cite{Giovannini2004,Kronberg1994,Widrow2003,Kandus2011,GrassoRubinstein,DurrerNeronov,Subramanian2016}. The upper and lower bounds on the 
strength of the present large-scale magnetic fields $B_{0}$ are 
given as $10^{-17} {\rm G} \lesssim B_{0} \lesssim 10^{-9} {\rm G}$ by the observations of the cosmic microwave background (CMB) 
\cite{Planck2015_PMF,Sutton2017} and the gamma rays from blazars \cite{NeronovVovk, Taylor, Tavecchio, Caprini}, respectively. The origin and 
evolution of these magnetic fields is a subject of intense studies.

Two classes of models for the origin of these magnetic fields are generally discussed (for a review, see Refs.~ 
\cite{Kandus2011,GrassoRubinstein,DurrerNeronov,Subramanian2016}). One possibility is that the observed fields result from the amplification 
during the structure formation of primordial magnetic fields produced in the early Universe. Another logical possibility is that the observed 
magnetic fields are of a purely astrophysical origin. The recent multifrequency blazar observation of large-scale magnetic fields in voids 
\cite{NeronovVovk, Taylor, Tavecchio, Caprini} with strength not less than $10^{-16}$ G \cite{Taylor} coherent on Mpc scale supports the case 
for primordial magnetic fields.

Various cosmological phase transitions could be considered as one of the possible ways of producing primordial magnetic fields 
\cite{Hogan1983,Quashnock1989,Vachaspati1991,Cheng1994,Sigl1997,Ahonen1998}. However, the comoving coherence length of such magnetic fields 
cannot be larger than the Hubble horizon at the phase transition, which is much smaller than Mpc today. Consequently, the most natural mechanism 
for the generation of the large-coherence-scale magnetic fields is inflation in the early Universe \cite{TurnerWidrow} through the exponential
stretching of wave modes during the accelerated expansion.

It is well known that quantum fluctuations of massless scalar and tensor fields are very strongly amplified in the inflationary stage and create 
considerable density inhomogeneities evolving later into the large scale structure of the observed
Universe \cite{Mukhanov1981,Hawking1982,Starobinsky1982,Guth1982,Bardeen1983} or relic gravitational waves
\cite{Grishchuk1975,Starobinsky1979,Rubakov1982}. However, since the Maxwell action is conformally invariant, the fluctuations in the 
electromagnetic field are not enhanced in the conformally flat expanding background of inflation \cite{Parker1968}. Therefore, in order to 
generate magnetic fields, one needs to break conformal invariance of the electromagnetic field, e.g. by coupling it to a scalar or a
pseudo-scalar field or to a curvature invariant. Although many ways to break the conformal invariance of the electromagnetic action during 
inflation were suggested in the literature \cite{Ratra,Dolgov1993,Gasperini1995,Giovannini2000,Atmjeet2014}, we adopt in our study the kinetic 
coupling model $f^{2}FF$ firstly introduced by Ratra \cite{Ratra}, where $f$ 
is a function of the inflaton field $\phi$ and $F$ is the electromagnetic field tensor. Depending on the form of the coupling $f$, this gives 
rise to different magnetic field power spectra \cite{Ratra,Giovannini2001,Bamba2004,Martin2008,Demozzi2009,Kanno2009,Ferreira}. Alternative models, in which magnetogenesis is 
driven by a rolling pseudo-Goldstone boson $\varphi$ through its coupling to the electromagnetic field in the form $\varphi F\tilde{F}$, are 
also very interesting and much studied \cite{Garretson1992,Anber2006,Durrer2011,Caprini2014,Fujita2015,Anber2015,Adshead2016}.

According to the most recent observational data by the Planck Collaboration \cite{Planck2015_infl}, the $R^{2}$ model proposed by Starobinsky in 
1980 \cite{Starobinsky} is the most favored among the models of inflation. For example, the chaotic inflationary models like large field 
inflation and natural inflation are disfavored due to their high tensor-to-scalar ratio. Therefore, we will study the inflationary 
magnetogenesis in the Starobinsky model in the present paper, which to the best of our knowledge was not previously investigated in
the literature. It is worth mentioning also that supergravity motivates a potential similar to
the Einstein gravity conformal representation of the $R^2$ inflationary model \cite{Ellis2013a,Ellis2013b,Buchmuller2013,Farakos2013,Ferrara2013}.

If the conformal invariance of the electromagnetic action is broken, the generation of cosmological magnetic fields can occur
after inflation before reheating, where the conductivity of the Universe becomes high, during the preheating stage \cite{Kobayashi}.
In such an epoch, the inflaton field oscillates around its potential minimum and the universe is effectively dominated by cold matter.  In dependence from
couplings between the inflaton and matter fields, this process sometimes might to proceed
non-perturbatively and   parametric resonance may play crucial role for bosonic fields 
\cite{KofmanLinde}, \cite{Rudenok:2014daa}. Since the electromagnetic field could be significantly amplified during preheating
\cite{Calzetta2002,Diaz-Gil2008,Jedamzik2010,Easther2011,Deskins2013,Kobayashi,Adshead2015,FujitaNamba}, we study also the postinflationary amplification of magnetic 
fields in the present paper and quantify how it affects the magnetic fields generated during inflation and preheating  in the Starobinsky model.

This paper is organized as follows. We  solve the background equations of Starobinsky model during inflation and preheating  in Sec.~\ref{section-starobinsky}. In Sec.~\ref{section-MF},  we consider the kinetic coupling of the inflaton field $\phi$ to the electromagnetic 
field, calculate the energy density of the generated
magnetic fields  and the evolution of the magnetic energy density through the subsequent stages of inflation and preheating is 
determined and analyzed. The summary of the obtained results is given in Sec.~\ref{section-conclusion}.

\section{Inflaton evolution during inflation and preheating}
\label{section-starobinsky}

Historically, one of the first models that exhibited inflation was the model suggested by Starobinsky \cite{Starobinsky} whose gravitational action reads as
\begin{equation}
\label{starobinsky}
S_{gr}=-\frac{M_{p}^{2}}{2}\int \!\!d^{4}x\,\sqrt{-g}\left[R-\frac{R^{2}}{6\mu^{2}}\right],
\end{equation}
where  $g={\rm det}\, g _{\lambda\nu},$  $g_{\lambda\nu}$ -- metric tensor, $\mu=1.3\cdot 10^{-5} M_{p}$ is a constant, which is fixed by the requirement to have the correct magnitude of the primordial 
perturbations \cite{Faulkner}  and 
$M_{p}=\frac{M_{Pl}}{\sqrt{8\pi}}=(8\pi G)^{-1/2}=2.4\cdot 10^{18}\ {\rm GeV}$ is the reduced Planck mass.   A conformal transformation $g_{\lambda\nu}\rightarrow \chi^{-1}g_{\lambda\nu}$ with 
$\chi=\exp\left(\sqrt{\frac{2}{3}}\frac{\phi}{M_{p}}\right)$
transforms the Lagrangian of theory (\ref{starobinsky}) into that of the usual Einstein gravity with a new spatially uniform scalar field $\phi$ (inflaton), whose 
potential reads
\begin{equation}
\label{infl-pot}
V(\phi)=\frac{3\mu^{2}}{4}\left(1-\exp\left[-\sqrt{\frac{2}{3}}\phi\right]\right)^{2}.
\end{equation}
It is quadratic in the vicinity of its minimum at $\phi=0$ and becomes flat at large values of $\phi$.

The time evolution of the Friedmann-Lema\^{\i}tre-Robertson-Walker (FLRW) universe with zero spatial curvature is described by the Friedmann equation
\begin{equation}
\label{friedmann}
H^{2}=\frac{1}{3 M_{p}^{2}}\left(\frac{1}{2}\dot{\phi}^{2}+V(\phi)\right),
\end{equation}
and the equation of motion for the  inflaton field in the FRW universe reads
\begin{equation}
\label{KGF}
\ddot{\phi}+3 H \dot{\phi}+\frac{\partial V}{\partial \phi}=0,
\end{equation}
where $H\equiv\frac{\dot{a}}{a}$ is the Hubble parameter, $a=a(t)$ is the scale factor. 
It is convenient to use the dimensionless quantities, where the inflaton field and Hubble parameter are expressed in Planck masses $M_{p}$, time in $M_{p}^{-1}$, and the Lagrangian density and inflaton effective potential in $M_{p}^{4}$

The Universe expands quasiexponentially during the inflation stage. In this regime the potential term dominates the kinetic one in 
Eq.~(\ref{friedmann}) and the ``friction'' term $3H\dot{\phi}$ dominates $\ddot{\phi}$ in Eq.~(\ref{KGF}). The slow roll parameters for the potential (\ref{infl-pot}) equal
\begin{equation}
\label{slow-roll-par-2}
\epsilon=\frac{1}{2}\left(\frac{V'}{V}\right)^{2}=\frac{4}{3\left(\exp\left[\sqrt{\frac{2}{3}}\phi\right]-1\right)^{2}},\quad \eta
=\frac{V''}{V}=\frac{4\left(2-\exp\left[\sqrt{\frac{2}{3}}\phi\right]\right)}{3\left(\exp\left[\sqrt{\frac{2}{3}}\phi\right]-1\right)^{2}},
\end{equation}
(here prime denotes derivatives with respect to the inflaton).
Inflation lasts until the slow-roll conditions are satisfied $\epsilon\ll 1,\ |\eta|\ll 1$. The duration of the inflation stage $\quad t_{inf}\simeq \frac{3}{2\mu}e^{\sqrt{\frac{2}{3}}\phi_{i}}$ depends on the initial value of the inflaton $\phi_{i}$. The minimal number of e-folds $N_{e}=\ln \frac{a_{e}}{a_{i}}\simeq \frac{3}{4}e^{\sqrt{\frac{2}{3}}\phi_{i}}$ necessary to solve the problems of horizon and flatness of the Universe is approximately $N_{min}\sim 60 - 70$. This implies $\phi_{i}=5.5$ and Eq.~(\ref{KGF}) in the slow roll approximation gives the initial value of the time derivative $\dot{\phi}_{i}=-\frac{V'(\phi_{i})}{\sqrt{3V(\phi_{i})}}$.

\begin{figure}
	\centering
	\includegraphics[scale=0.48]{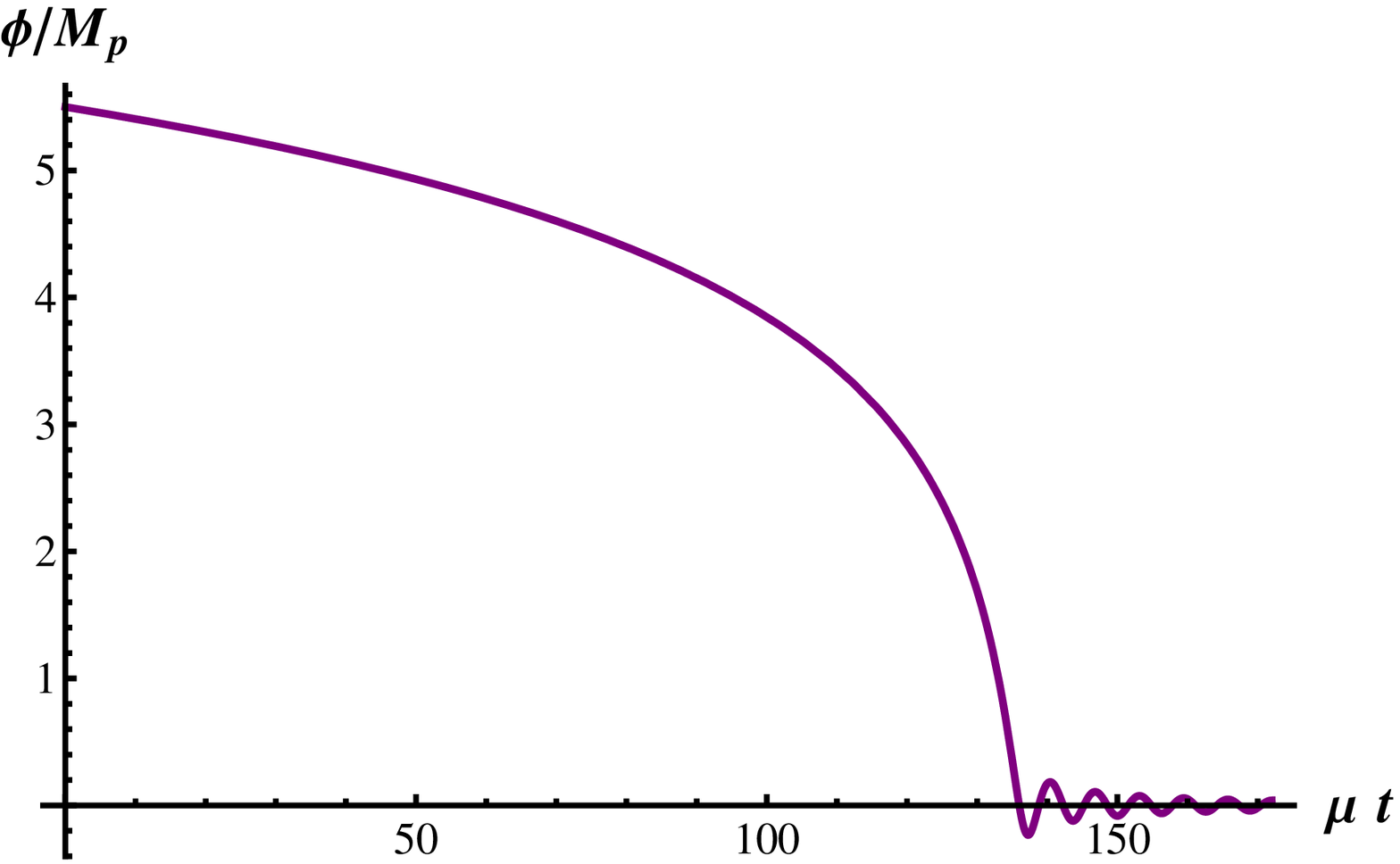} \hfill
	\includegraphics[scale=0.48]{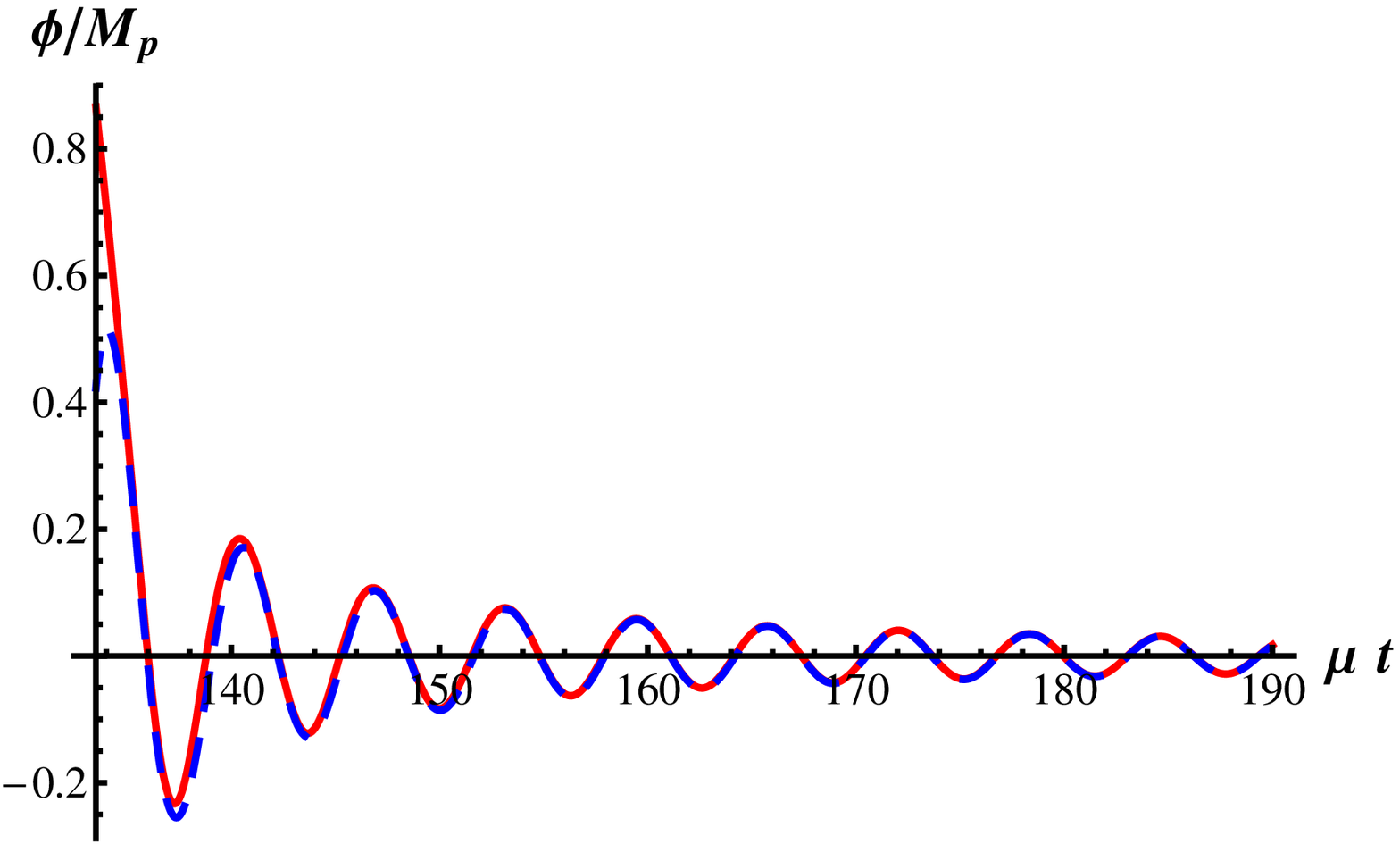}
	\caption{Left panel: The time dependence of the inflaton field during inflation and after it. Right panel: Oscillations of the 
	inflaton field after the end of inflation (red solid line) and the approximate function (blue dashed line).
	\label{inflaton}}
\end{figure}

To determine the time dependence of the inflaton field and the scale factor, Eqs.~(\ref{friedmann})-(\ref{KGF}) were integrated with the initial conditions, discussed above, and the following approximate solution has been obtained by using the slow roll conditions:
\begin{eqnarray}
a(t)&=&\exp\left(\frac{\mu t}{2}\right)\left[1-\frac{t}{t_{inf}}\right]^{3/4},\ \ H(t)=\frac{\mu}{2}\left(1-\frac{1}{\exp\left(\sqrt{\frac{2}{3}}\phi_{i}\right)-\frac{2}{3}\mu t}\right),\\
\label{time-dep-infl}
\phi(t)&=&\phi_{i}+\sqrt{\frac{3}{2}}{\rm ln}\left[1-\frac{t}{t_{inf}}\right].
\end{eqnarray}
This solution satisfactory fits the numerical solution to Eqs.~(\ref{friedmann})-(\ref{KGF}) for $\mu(t_{inf}-t)\gg 1$. The time dependence of the inflaton is plotted in the left panel in Fig.~\ref{inflaton}.

The inflaton field during preheating after the end of inflation behaves like a dust (pressure $p=0$), therefore, $a\sim (t-t_{s})^{2/3}$, where $t_{s}$ is the moment of time when the Universe became dust-dominated. Since the amplitude of fast oscillations decreases in time as $\sim a^{-3/2}$, the approximate solution  during the preheating stage is $\phi(t)=\frac{C}{\mu(t-t_{s})}\sin\left[\mu\left(t-t_{0}\right)\right]$,
where $C=\sqrt{8/3}$ were found taking into account that  in the vicinity of the minimum the inflaton potential (\ref{infl-pot}) behaves like a  parabola $V(\phi)\approx \frac{\mu^{2}\phi^{2}}{2}$
and the Hubble parameter equals $H=\frac{2}{3(t-t_{s})}$ during preheating. 
The phase shift time $t_{0}$ and $t_{s}$ could be determined numerically from the best fit condition. We found $t_{s}=1.008\cdot 10^{7}$ and $t_{0}=1.0216\cdot 10^{7}$.
Introducing the shifted time $\tau=t-t_{s}$, one obtain
\begin{eqnarray}
\label{scale-preh}
a(\tau)&=&a_{e}\left(\frac{\tau}{\tau_{0}}\right)^{2/3},\qquad H(\tau)=\frac{2}{3\tau},\\
\label{inflaton-oscillations}
\phi(\tau)&=&\frac{\sqrt{8/3}}{\mu\tau}\sin[\mu(\tau-\tau_{0})],
\end{eqnarray}
where $\mu\tau_{0}=1.77$, $a_{e}$ is the value of the scale factor at the end of inflation.
The approximate solution (\ref{inflaton-oscillations}) is showed in the right panel in Fig.~\ref{inflaton} by the blue dashed line. 
Obviously, it fits the accurate solution (red solid line) very well.

\section{Magnetic field generation}
\label{section-MF}

In order to study magnetogenesis in the Starobinsky model, we consider the kinetic coupling of the inflaton field $\phi$ to the electromagnetic 
field $F_{\lambda\nu}\equiv \partial_\lambda A_\nu-\partial_\nu  A_\lambda,$ characterized by its four-vector potential $A_\nu$
\begin{equation}
\label{lagr-dens}
S_{int}=\int \!\!d^{4}x\,\sqrt{-g} \,\mathcal{L}_{int}, \qquad \mathcal{L}_{int}=-\frac{f^{2}(\phi)}{4}F_{\lambda\nu}F^{\lambda\nu}, 
\end{equation}
where $F^{\lambda\nu}=g^{\lambda\gamma}g^{\beta\nu}F_{\gamma\beta},$  and $f(\phi)$ is coupling function, introduced by Ratra \cite{Ratra}
\begin{equation}
\label{coupling-function}
f(\phi)=\exp(\alpha\phi).
\end{equation}
This function with free parameter $\alpha$  gives the correct value $f=1$ after the end of preheating and, consequently, the correct value of the electron charge today. In 
addition, this function does not cause the strong coupling problem during the inflation stage, where $\phi\gg1$.

The equation of motion for the electromagnetic vector potential in the Coulomb gauge $A_{0}=0$, $\partial_{i}A^{i}=0$ has the form
\begin{equation}
\label{el-mag-1}
\ddot{A}_{i}(t,x)+\left(H+2\frac{\dot{f}}{f}\right)\dot{A}_{i}(t,x)-\partial_{j}\partial^{j}A_{i}(t,x)=0.
\end{equation}

Quantizing the electromagnetic field, the vector potential can be decomposed into the sum over creation $\hat{b}^{\dagger}_{k,\lambda}$ and annihilation operators $\hat{b}_{k,\lambda}$:
\begin{equation}
\label{Fourier}
\hat{A}_{j}(t,x)=\int\frac{d^{3}k}{(2\pi)^{2/3}}\sum_{\lambda=1}^{2}\left[\epsilon^{\lambda}_{j}(k)\hat{b}_{k,\lambda}A(t,k)e^{i k\cdot x}
+\epsilon^{*\lambda}_{j}(k)\hat{b}^{\dagger}_{k,\lambda}A^{*}(t,k)e^{-i k\cdot x}\right],
\end{equation}
where $\epsilon^{\lambda}_{j}(k)$, $\lambda=1,\,2$ are two independent transverse polarization vectors. The time evolution of the function $\mathcal{A}(t,k)=f(t)a(t)A(t,k)$ is governed by the equation
\begin{equation}
\label{modes-evol-real}
\ddot{\mathcal{A}}(t,k)+H\dot{\mathcal{A}}(t,k)+\left(\frac{k^{2}}{a^{2}(t)}-H\frac{\dot{f}}{f}-\frac{\ddot{f}}{f}\right)\mathcal{A}(t,k)=0.
\end{equation}
The scale factor $a(t)$ in the definition of $\mathcal{A}$ originates from the presence of the polarization vectors $\epsilon^{\lambda}_{j}$ in the Fourier expansion (\ref{Fourier}), which contain the scale factor in their explicit expressions \cite{Martin2008}. 

One rewrite this equation  in conformal time $\eta(t)=\int^{t}\!\frac{dt'}{a(t')}.$

\begin{equation}
\label{modes-evol-conf}
\mathcal{A}''(\eta,k)+\left[k^{2}-\frac{f''}{f}\right]\mathcal{A}(\eta,k)=0.
\end{equation}
(prime denotes derivatives with respect to the conformal time)
we can determine 
the initial condition to Eq.~(\ref{modes-evol-conf})  from the asymptotic behavior in the early 
stages, where we assume $f=f(\phi_{i})={\rm const}$,
\begin{equation}
\label{boundary-condition}
\mathcal{A}\simeq\mathcal{A}_{free}(\eta,k)=\frac{1}{\sqrt{2k}}e^{-ik\eta}, \qquad -k\eta\gg 1.
\end{equation}

Covariantly defined electric and magnetic fields seen by an observer
characterized by the 4-velocity vector $u^{\mu}$ have the following form \cite{Barrow}:
\begin{equation}
E_{\mu}=u^{\nu}F_{\mu\nu},\qquad B_{\mu}=\frac{1}{2}\eta_{\mu\nu\rho\sigma}F^{\nu\rho}u^{\sigma},
\end{equation}
where $\eta_{\mu\nu\rho\sigma}$ is the totally antisymmetric tensor with $\eta_{0123}=\sqrt{-g}$. For the comoving observer with
$u^{\mu}=(1,\vec{0})$, we find in the Coulomb gauge
\begin{equation}
E_{i}=-\partial_{t}A_{i}, \qquad B_{i}=\frac{1}{a}\epsilon_{ijk}\partial_{j}A_{k}
\end{equation}
with $\epsilon_{123}=1$.

The interaction Lagrangian (\ref{lagr-dens}) makes the following contribution to the stress-energy tensor:
\begin{equation}
T_{\lambda\nu}=-\frac{2}{\sqrt{-g}}\frac{\delta S_{int}}{\delta g^{\lambda\nu}}=-f^{2}(\phi)g^{\gamma\beta}F_{\lambda\gamma}F_{\beta\nu}
+g_{\lambda\nu}\mathcal{L}_{int}.
\end{equation}

Then, we could determine the energy density of the electromagnetic field as $\rho=-\langle T^0_0\rangle$. The 'magnetic' part of the energy density $\rho_{B}$ does not contain time derivatives of the vector potential $\partial_{0}A_{i}$, while only the 'electric' part $\rho_{E}$ contains such derivatives. They equal \cite{Martin2008}
\begin{eqnarray}
\rho_{B}(t)&=&\int_{0}^{+\infty}\!\frac{dk}{k}\frac{d\rho_{B}(t,k)}{d\ln k}=\frac{1}{2\pi^{2}}\int_{0}^{+\infty}\!\frac{dk}{k}\left(\frac{k}{a(t)}\right)^{4}k|\mathcal{A}(t,k)|^{2},\\
\rho_{E}(t)&=&\int_{0}^{+\infty}\!\frac{dk}{k}\frac{d\rho_{E}(t,k)}{d\ln k}=\frac{1}{2\pi^{2}}\int_{0}^{+\infty}\!\frac{dk}{k}\left(\frac{k}{a(t)}\right)^{2}k\,f^{2}(t)\left|\frac{\partial}{\partial t}\left(\frac{\mathcal{A}(t,k)}{f(t)}\right)\right|^{2}.
\end{eqnarray}

Numerically solving Eq.~(\ref{modes-evol-real}) with the corresponding boundary conditions, we determine the vector 
potential $\mathcal{A}(t,k)$ and the power spectrum of generated magnetic and electric fields, $\frac{d\rho_{B}(t,k)}{d\ln k}$ and $\frac{d\rho_{E}(t,k)}{d\ln k}$.

Once we obtained the spectrum of the magnetic field, we should rescale it up to present time. For this, we should determine the value of 
the scale factor at the present time compared to its value the at end of preheating
\begin{equation}
\frac{a_{0}}{a_{e}}\sim\frac{T_{max}}{T_{0}}.
\end{equation}
If we take $T_{max}\sim 10^{15} \ {\rm GeV}$ and $T_{0}=2.3\cdot 10^{-13}\ {\rm GeV}$, we obtain
\begin{equation}
\frac{a_{0}}{a_{e}}\sim 10^{28}.
\end{equation}

The authors of Ref.~[\onlinecite{GrassoRubinstein}] define the so-called cosmic diffusion length as the minimal size of a magnetic configuration 
which can survive diffusion during the Universe lifetime and estimate it as $r_{diff}\sim 1\ {\rm A.U.}=1.5\cdot 10^{13}\ {\rm cm}$. The 
corresponding wave vector in our epoch is $k_{diff}/a_{0}\sim 1\ {\rm A.U.}^{-1}\approx 1.3\cdot 10^{-27}\ {\rm GeV}$. Therefore, in what 
follows, we will be interested only in modes with $k<k_{diff}$.

\subsection{Magnetogenesis during inflation}
\label{mag-gen-infl}

It is interesting to follow the time evolution of the amplitude of a given mode. When the mode is below the horizon it oscillates in 
time without significant changes of its amplitude. After the mode crosses the horizon the amplitude begin to diminish due to the Universe 
expansion. Therefore, the earlier mode exits the horizon, the smaller amplitude it has at the end. This process lasts nearly to the end of 
inflation, when the evolution of the inflaton field starts to deviate from the slow-rolling regime. Let us rewrite Eq.~(\ref{modes-evol-real}) in terms of a rescaled field $\mathcal{F}(t,k)=a^{1/2}(t)\mathcal{A}(t,k)$:
\begin{equation}
\ddot{\mathcal{F}}(k,t)+\omega_{k}^{2}(t)\mathcal{F}(k,\tau)=0, \qquad \omega_{k}^{2}(t)=\frac{k^{2}}{a^{2}(t)}-H\frac{\dot{f}}{f}-\frac{\ddot{f}}{f}+\frac{1}{4}H^{2}
-\frac{1}{2}\frac{\ddot{a}}{a}.
\label{F-equation}
\end{equation}
Taking into account the explicit form of the coupling function (\ref{coupling-function}), we can represent $\omega_{k}^{2}$ as follows:
\begin{equation}
\label{freq}
\omega_{k}^{2}(t)=\frac{k^{2}}{a^{2}(t)}-H\alpha\dot{\phi}-\alpha^{2}\dot{\phi}^{2}-\alpha\ddot{\phi}+\frac{1}{4}H^{2}
-\frac{1}{2}\frac{\ddot{a}}{a}.
\end{equation}
Near the end of inflation, the inflaton field changes more quickly and for $\alpha\gg1$ the term $-\alpha^{2}\dot{\phi}^{2}$ begins 
to dominate the rest of the terms in Eq.~(\ref{freq}). This happens roughly 10 e-folds before the end of inflation. Then function 
(\ref{freq}) changes its sign to negative and an instability occurs. As a result, all modes beyond the horizon undergo amplification. We 
would like to emphasize that such an amplification is a built-in property of the model under consideration. Indeed, at first, the growth by
modulus of the time derivative of the inflaton field near the end of inflation is connected with the form of the effective potential
(\ref{infl-pot}) in the Starobinsky model. At second, the presence of the term $-\alpha^{2}\dot{\phi}^{2}$ in Eq.~(\ref{freq}) is due to the
exponential form of the kinetic coupling function. Finally, large values $\alpha\gg 1$ ensure the dominance of this term.

\begin{figure}
	\centering
	\includegraphics[scale=0.6]{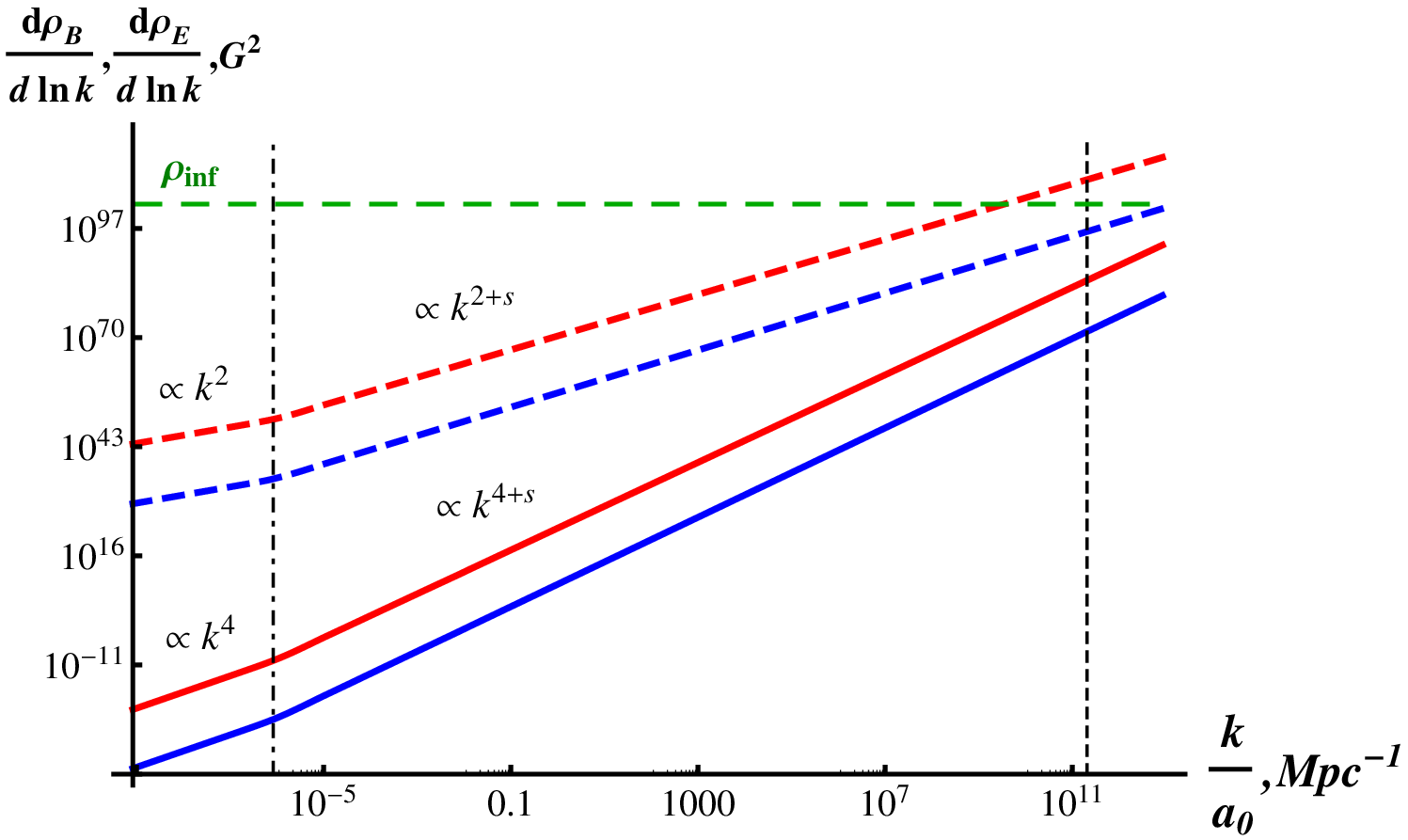}
	\includegraphics[scale=0.38]{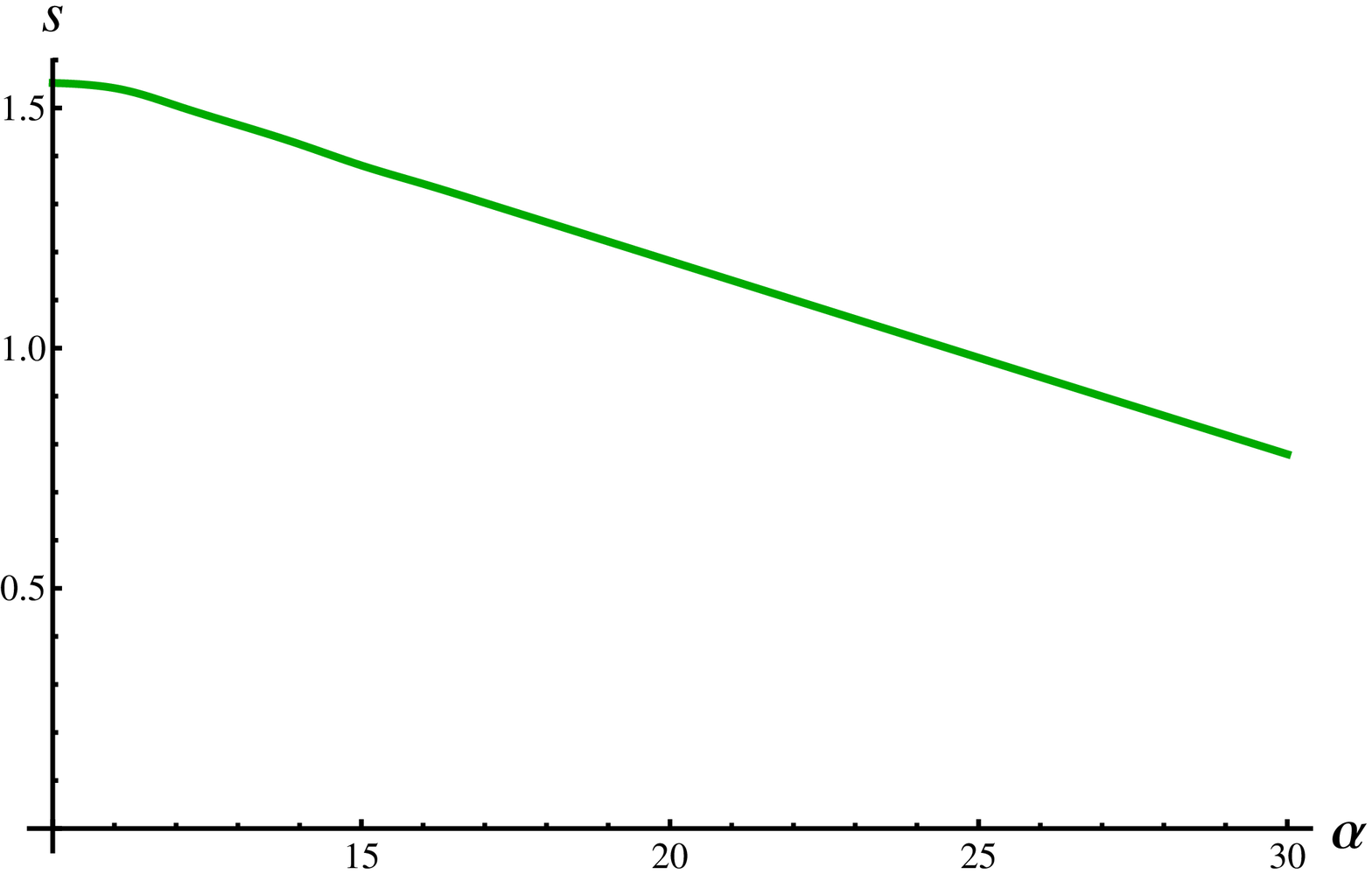}
	\caption{Left panel: The power spectrum of magnetic (solid lines) and electric (dashed lines) fields generated during inflation for $\alpha=12$ (blue lines) and $\alpha=15$ (red lines). The vertical dash-dotted line separates the modes which were beyond the horizon even at the beginning of the inflation. The vertical dashed line shows the last mode $k/a_{0}\sim 1\ {\rm A.U.}$ which survives diffusion during the evolution of the Universe. The horizontal dashed line shows the energy density of the inflaton field $\rho_{inf}$ during the inflation. Right panel: the dependence of the anomalous slope $s$ on the parameter $\alpha$.
	\label{spectr_infl}}
\end{figure}

We plot the power spectrum of magnetic and electric fields at the end of inflation in Fig.~\ref{spectr_infl} for two different values of 
parameter $\alpha$. Clearly, these spectra have similar behavior. For modes, which are beyond the horizon even at the beginning of the inflation 
$k\ll \mu/2$, i.e. $k/a_{0}\ll 4\cdot 10^{-6} \ {\rm Mpc^{-1}}$ (to the left from the dash-dotted vertical line), the magnetic spectrum behaves 
like $\propto k^{4}$. For modes, which exit the horizon during inflation, the spectrum is more steep, $\propto k^{4+s}$, where $s$ is the 
anomalous slope. This behavior is expected to be true up to momenta $k\sim a_{inf}H_{inf}$, which exit the horizon at the end of inflation. For 
larger momenta, the corresponding modes never exit the horizon. Therefore, these modes oscillate in time during the whole inflation 
stage and undergo neither diminishing nor amplification. As a result, their spectrum remains unchanged and behaves like
$\propto k^{4}$. We are interested among all modes only in those with momenta up to $k/a_{0}\sim 1\ {\rm A.U.}$ (dashed 
vertical line), because shorter waves would not survive the cosmic diffusion \cite{GrassoRubinstein}. Therefore, we do not show the spectrum for 
larger momenta. 

The power spectrum of the electric fields, generated during inflation, shows much more larger values and scales like $\propto k^{2}$ for modes, which are beyond the horizon even at the beginning of the inflation, and $\propto k^{2+s}$ for modes, which exit the horizon during inflation. Should be  noted, that the anomalous slope $s$ is the same as for magnetic field power spectrum. The right panel in Fig.~\ref{spectr_infl} shows, that the dependence of the anomalous slope $s$ on $\alpha$ is linear and can be approximated as
\begin{equation}
s(\alpha)\approx 1.98-0.04 \alpha.
\end{equation}

Let us estimate whether the back-reaction problem occurs. According to Refs.~[\onlinecite{Martin2008,Fujita2012}], the model is free of this difficulty when the following condition is satisfied for all modes with $k < k_{diff}$:
\begin{equation}
\label{back-react}
\left.\frac{d\rho_{E}(t,k)}{d\ln k}\right|_{end}+\left.\frac{d\rho_{B}(t,k)}{d\ln k}\right|_{end}<\rho_{inf},
\end{equation}
where $\rho_{inf}$ is the energy density of the inflaton field during inflation. In our case it could be estimated as:
\begin{equation}
\rho_{inf}=3H_{inf}^{2}M_{p}^{2}\approx \frac{3}{4}\mu^{2}M_{p}^{2}\sim 10^{103} {\rm Gauss}^{2}.
\end{equation}
It is shown in Fig.~\ref{spectr_infl} by a horizontal green dashed line. This figure implies, that the back-reaction problem does not occur for all cosmologically relevant modes for $\alpha=12$. As for $\alpha=15$, condition (\ref{back-react}) is violated only near $k_{diff}$. All calculations were done assuming, that $a_{0}/a_{e}\sim 10^{28}$, which corresponds to $T_{max}\sim 10^{15}\, {\rm GeV}$. The 'electric' part gives the leading contribution to the back-reaction and it scales like $\propto\left(a_{0}/a_{e}\right)^{2+s}$. Therefore, if the maximal temperature during preheating were lower, then, the ratio $a_{0}/a_{e}$ would be less and the comoving wave number, which corresponds to present cosmic diffusion scale of 1 A.U., $k_{diff}$, would be lower. As a result, the back-reaction problem would be ameliorated.

As it was discussed in Ref.~[\onlinecite{Martin2008}], during reheating, the conductivity jumps and, as a consequence, the electric field vanishes. Thus, if one checks that, at the end of inflation, the electric field can not cause a back-reaction problem, then we are guaranteed that the complete scenario is consistent. As we will see in the next subsection, the stage of preheating also contributes to the generated spectrum, but for $\alpha=12-15$ this contribution is only a few orders of magnitude, therefore, it cannot cause the back-reaction problem, if it was not caused during inflation.

\subsection{Magnetogenesis during preheating}

In the previous subsection, we found the power spectrum of magnetic field generated during the inflation stage. After inflation the scalar field 
oscillates in the vicinity of the minimum of its potential, producing various particles. Important to study how these fast 
oscillations affect the power spectrum of magnetic fields obtained earlier. For this purpose we define the transfer function
$\mathcal{T}(k;\tau,\tau_{0})$, which shows the relative enhancement of a given mode $k$ at the moment of time $\tau$ compared to the moment of
the beginning of the preheating stage $\tau_{0}$
\begin{equation}
\mathcal{T}(k;\tau,\tau_{0})=\frac{|\mathcal{A}(k,\tau)|^{2}}{|\mathcal{A}(k,\tau_{0})|^{2}},
\end{equation}
obtain the power spectrum at the end of preheating stage and rescale it until the present time
\begin{equation}
\left.\frac{d\rho_{B}}{d\ln k}\right|_{\rm now}=\left.\frac{d\rho_{B}}{d\ln k}\right|_{inf}\cdot\mathcal{T}(k;\tau_{e},\tau_{0})
\frac{a_{e}^{4}}{a_{0}^{4}},
\end{equation}
where $\tau_{e}$ is the time of the end of the preheating stage, when the amplitude stops increasing.

To obtain the transfer-function we have to solve Eq.~(\ref{modes-evol-real}). For the rescaled field
$\mathcal{F}(k,\tau)=a^{1/2}(\tau) \mathcal{A}(k,\tau)$ [here $\tau$ is a shifted time, defined before Eq.~(\ref{scale-preh})], we obtain Eq.~(\ref{F-equation}), which, taking into account the coupling function (\ref{coupling-function}), Eq.~(\ref{inflaton-oscillations}), and retaining the monotonous and the largest oscillating terms in the brackets, could be brought to the Mathieu-like equation by the change of variable $\mu(\tau-\tau_{0})=2z-\pi/2:$
\begin{equation}
\mathcal{F}''(z)+[a_{M}(k,z)-2q_{M}(z)\cos(2z)]\mathcal{F}(z)=0,
\end{equation}
where $a_{M}(k,z)=\frac{4k^{2}}{\mu^{2}a^{2}(z)}+\frac{8}{9(2z-\pi/2+\mu\tau_{0})^{2}}$ and $q_{M}(z)=\frac{4\alpha\sqrt{2}}{\sqrt{3}(2z-\pi/2+\mu\tau_{0})}$ are the monotonously decreasing functions of $z$. For constant and sufficiently large $q_{M}$ the Mathieu equation has exponentially growing solutions. This is the parametric resonance situation. However, $q_{M}(z)$ decreases with time and the system exits the resonance band and, as a result, the exponential growth stops. Therefore, the most  considerable enhancement takes place during the first few oscillations of the inflaton.

\begin{figure}
	\centering
	\includegraphics[scale=0.4]{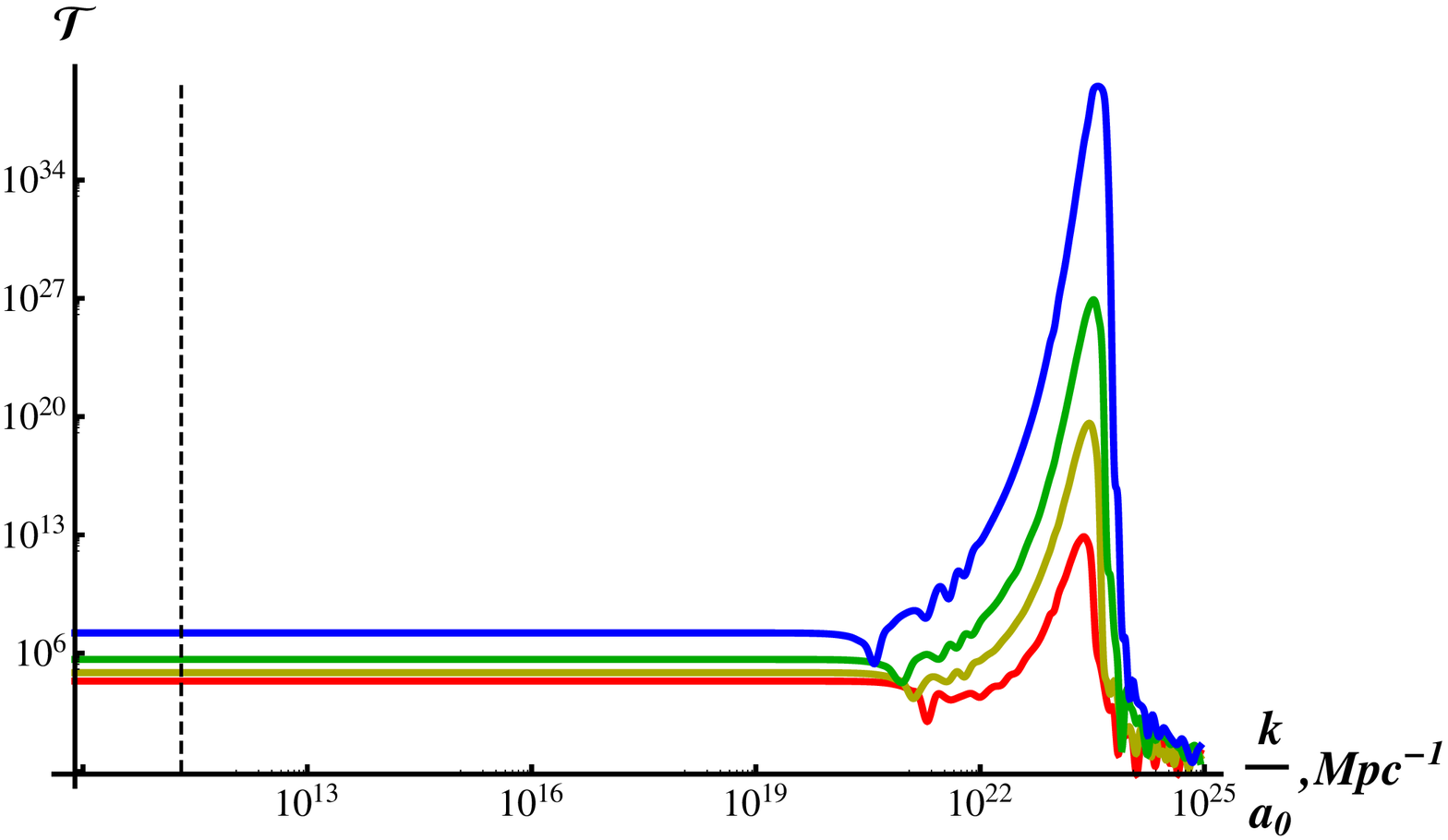}
	\includegraphics[scale=0.4]{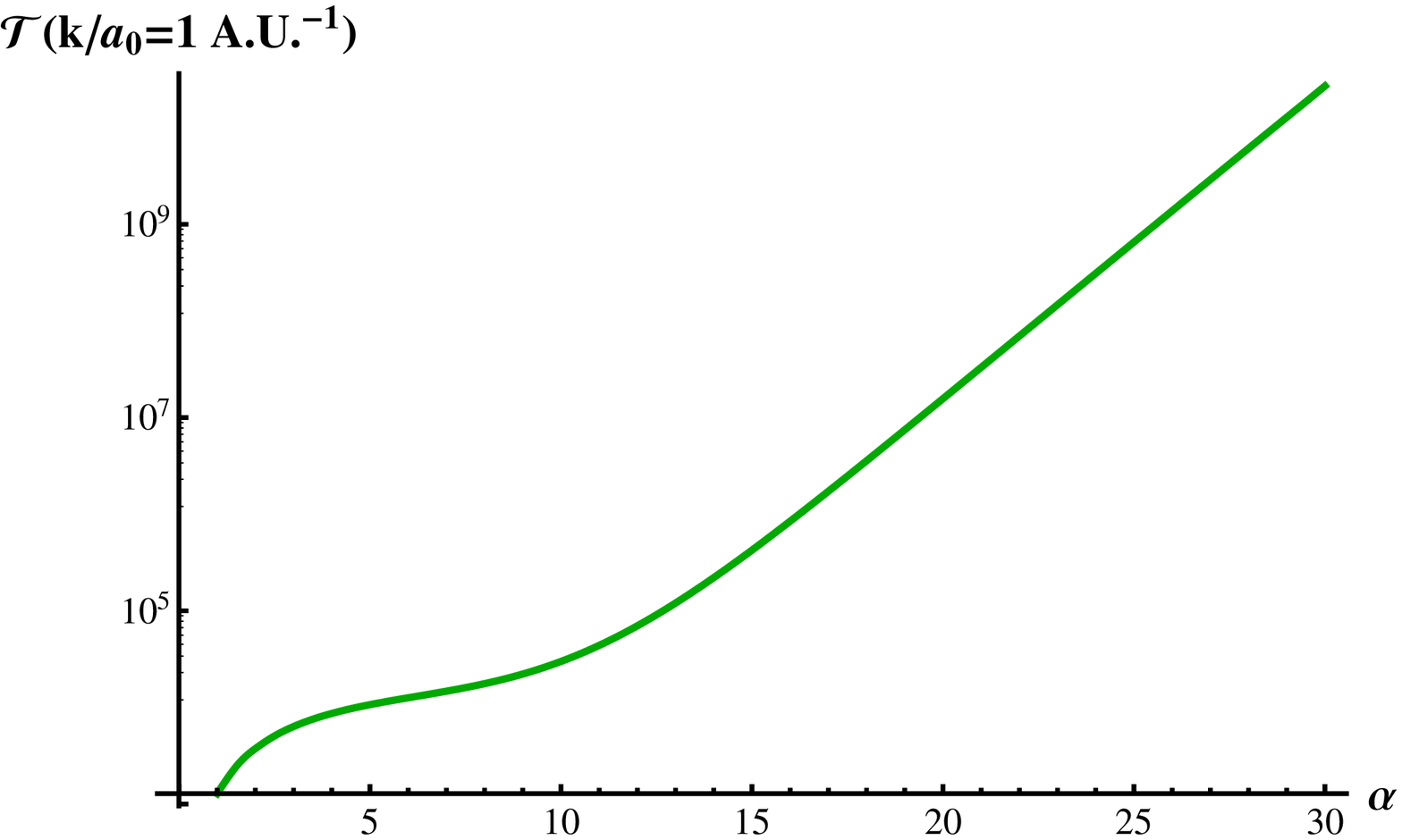}
	\caption{Left panel: The transfer-function of the power spectrum at the end of preheating stage for $\alpha=9$ (red line), $\alpha=12$ 
	(yellow line),
	$\alpha=15$ (green line), and $\alpha=20$ (blue line). The vertical dashed line shows the last mode which survives the cosmic diffusion 
	until present time. Right panel: The transfer-function for the mode with momentum $k/a_{0}= 1\ {\rm A.U.}^{-1}$ as a function of the
	parameter $\alpha$.
	\label{spectr-preh}}
\end{figure}

The transfer-function of the power spectrum fixed at $\mu\tau_{e}=1000$ (at the end of preheating stage) is plotted in the left 
panel of Fig.~\ref{spectr-preh}. Obviously, for $k\ll a_{inf}H_{inf}$, i.e. $k/a_{0}\ll 10^{21}\ {\rm Mpc}^{-1}$, the transfer-function is constant. 
For modes, which re-entered the horizon  $k\gtrsim a_{inf}H_{inf}$, the transfer-function demonstrates an oscillatory behavior and its amplitude 
grows faster. At very large momenta, $k/a_{0}\gtrsim 10^{24}\ {\rm Mpc}^{-1}$, enhancement is absent and we observe the original spectrum. The 
picture is qualitatively similar to Fig.~1d in Ref.[\onlinecite{Kobayashi}]. As it was mentioned above, we are interested in modes with momenta 
$k/a_{0}\lesssim 1\ {\rm A.U.}$ For these modes, we can consider the transfer-function as a constant. Its dependence on $\alpha$ is shown in the 
right panel of Fig.~\ref{spectr-preh}.

\begin{figure}
	\centering
	\includegraphics[scale=0.4]{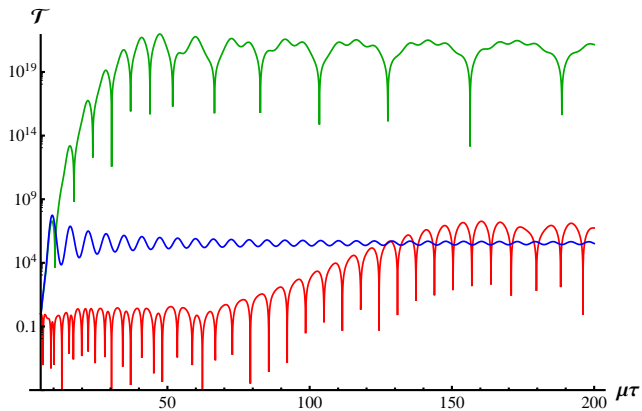}
	\caption{The time dependence of the transfer-function during the preheating stage for three modes
	$k/a_{0}=2\cdot 10^{11} {\rm Mpc}^{-1}$ (blue line),  $k/a_{0}=1.5\cdot 10^{23} {\rm Mpc}^{-1}$ (green line), and $k/a_{0}=7\cdot 10^{23} {\rm Mpc}^{-1}$ (red line).
	\label{par-res}}
\end{figure}

There are some interesting features in the time evolution of the transfer-function during preheating, see Fig.~\ref{par-res}. For the modes with small momenta, which can survive diffusion, $k/a_{0}< 1\ {\rm A.U.}^{-1}$, parametric resonance is rather inefficient and an amplification occurs only during the first oscillation of the inflaton (the blue line in Fig.~\ref{par-res}). The situation changes for the modes with the physical momentum at the beginning of preheating $k_{ph}=k/a(\tau_{0})$, which is comparable with the frequency of the inflaton oscillations $\mu$. For these modes parametric resonance is more efficient and the amplitude grows during the first 5--10 oscillations, see the green line in Fig.~\ref{par-res}. This causes the peak in the spectrum, see the left panel in Fig.~\ref{spectr-preh}.  However, for larger momenta, the resonance does not occur at the beginning of preheating stage and a stochastic behavior is observed (the red line on Fig.~\ref{par-res} for $\mu\tau<60$). When the Universe expansion red-shifts the physical momentum to the values comparable with $\mu$, the resonance turns on and we observe the amplification of the amplitude during a few oscillations (the red line on Fig.~\ref{par-res} for $60<\mu\tau<150$). This amplification is not as large as for the green line, because at the moment, when it starts, the value of the parameter $q_{M}$ is smaller due to the Universe expansion. This explains the decreasing 'tail' of the spectrum in Fig.~\ref{spectr-preh}. Although these modes demonstrate a rather interesting behavior, they would not survive the diffusion during further evolution of the Universe, therefore, they do not contribute to the present value of the magnetic field.

\begin{figure}
	\centering
	\includegraphics[scale=0.4]{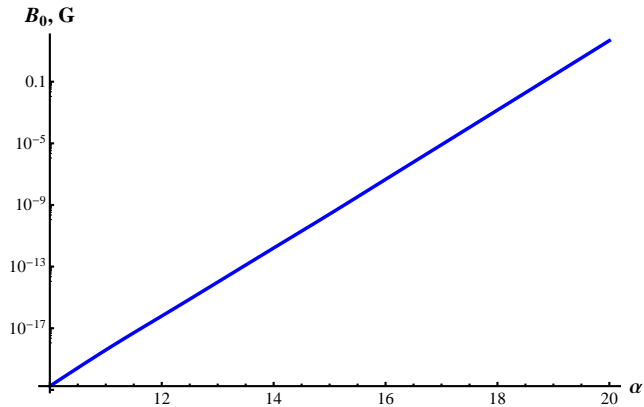}
	\caption{The generated magnetic field as a function of $\alpha$.
		\label{MF}}
\end{figure}

Taking into account all previous results, we can now calculate the generated magnetic field:

\begin{equation}
B_{0}=\sqrt{2\rho_{B}}=\sqrt{2\int_{0}^{k_{diff}}\frac{dk}{k}\frac{d\rho_{B}}{d \ln k}}.
\end{equation}
The strength of this magnetic field strongly depends on $\alpha$. The corresponding dependence is plotted in Fig.~\ref{MF}. This figure implies
that the magnetic fields which correspond to the present observations could be obtained at $\alpha\sim 12-15$.

\section{Conclusion}
\label{section-conclusion}

In this work, we studied the generation of large scale magnetic fields in the Starobinsky model of inflation, which is favored by 
the latest results of Planck 2015 \cite{Planck2015_infl}. In order to break the conformal invariance of the electromagnetic action, we chose
the kinetic coupling $f^{2}(\phi)FF$ of the inflaton field with the electromagnetic field through the exponential coupling function 
$f(\phi)=\exp(\alpha\phi)$, which does not cause the strong coupling problem during inflation. To the best of our knowledge, this form of 
coupling function in combination with the Starobinsky model of inflation has not been considered in the literature before.

In addition, we examined the possibility of further amplification of generated magnetic field during the preheating stage, when the inflaton 
oscillates in the vicinity of the minimum of its potential. During this stage, the Universe is effectively matter-dominated, the inflaton's 
amplitude of oscillations decreases in time. This reduces the effectiveness of the parametric resonance and gives the exponential amplification of magnetic field  during the first few oscillations of the inflaton field.

We found that it is possible in such a model to generate the large scale magnetic fields with strength $\gtrsim 10^{-15}$~Gauss at the present epoch during the inflation and preheating stages in a certain range of the parameter $\alpha\sim 12 - 15$. The spectrum of generated magnetic fields is blue with the spectral index $n=1+s$, $s>0$. When a certain mode exits the horizon, its amplitude diminishes due to the Universe expansion. Therefore, the earlier mode exits the horizon, the smaller amplitude it has in the end. This explains the blue spectrum of generated magnetic fields. It is necessary to emphasize that near the end of inflation the evolution of the  inflaton field starts to deviate from the slow-rolling regime and leads to a significant increase by modulus of the time derivative of the inflaton field. As we showed in Subsec.~\ref{mag-gen-infl}, this causes an instability in the equation governing the evolution of the electromagnetic field. As a result, all modes beyond the horizon undergo amplification. This is a consequence of the three independent features of our model: (i) the effective potential of the Starobinsky model (\ref{infl-pot}) causes the growth of $|\dot{\phi}|$ near the end of inflation, (ii) the exponential form of the kinetic coupling function produces the term $\alpha^{2}\dot{\phi}^{2}$ in Eq.~(\ref{modes-evol-real}), (iii) large values $\alpha\gg 1$ lead to the domination of this term and to an instability. Of course, the modes, which do not exit the horizon until the end of inflation undergo neither diminishing nor amplification and remain unchanged.

According to Refs.~[\onlinecite{FujitaNamba,Demozzi2009,Ferreira}], the kinetic coupling model often faces the back-reaction problem. 
In our model, the kinetic coupling function does not scale like $f\propto a^{\alpha}$ during the inflation stage and, therefore, the 
back-reaction could not be treated by the methods considered in Ref.~[\onlinecite{Demozzi2009}]. Therefore, we used a numerical analysis
and found that for a certain range of values of the coupling parameter, $\alpha=12-15$, our model avoids the back-reaction problem for 
all relevant modes. Other constraints on inflationary magnetogenesis are often enforced by the requirement that the back-reaction of generated 
magnetic fields on the evolution of primordial curvature perturbations is small \cite{Ferreira}. We plan to address this issue elsewhere.

We found also that the value of the generated magnetic field scales with the maximal temperature during preheating as
$B_{0}\propto T_{max}^{s/2}$. Therefore, a lower maximal temperature $T_{max}$ would give a lower value of $B_{0}$ for a given $\alpha$. On 
the other hand, the energy density of electromagnetic fields at the end of inflation, which can cause the back-reaction, scales like
$\propto T_{max}^{2+s}$ and decreases much faster compared to $B_0$ as $T_{max}$ decreases. Therefore, lower values of $T_{max}$ make possible
to extend the range of possible values of $\alpha$.

Finally, we would like to mention that it would be interesting to extend our study by taking into account the role of chiral anomaly
\cite{Joyce} and helicity \cite{Cornwall} on the evolution of magnetic fields in the early Universe. According to Ref.~\cite{Boyarsky}, the
inclusion of anomalous currents leads to an inverse cascade, where a part of the energy of magnetic fields is transferred from shorter to longer 
wavelengths and, thus, escape the dissipation during the evolution of the Universe (for a recent discussion, see Ref.~\cite{Shtanov2016}).  
The role of inhomogeneities in primordial chiral plasma was addressed in Refs.\cite{Boyarsky-hydrodynamics, Gorbar:2016qfh}. By numerically
studying the anomalous Maxwell equations, it was shown \cite{ Gorbar:2016klv} that due to the effects of diffusion these inhomogeneities do not 
prevent the anomaly-driven inverse cascade. On the other hand, it was shown long time ago \cite{Pouquet} that the inverse cascade is driven by
helical magnetic turbulence. Therefore, it is interesting and urgent to study the magnetohydrodynamics of the primordial plasma accounting for 
the effects of both chirality and turbulence. A first step in such an analysis was recently done in Ref.~\cite{Dvornikov}.

\vspace{5mm}

\begin{acknowledgments}
The work of S.~V., E.~G. and I.~R. is supported partially by the Swiss National Science Foundation, Grant No. SCOPE IZ7370-152581. S.~V. is grateful to the Swiss 
National Science Foundation (individual Grant No. IZKOZ2-154984).
S.~V.\@ is grateful to Prof.\@ Marc Vanderhaeghen and Dr.\@ Vladimir
Pascalutsa for their support and kind hospitality at the Institut f\"{u}r Kernphysik,
Johannes Gutenberg-Universit\"{a}t Mainz, Germany, where the part of this work was done. O.~S. is grateful to Prof.~Alexei Boyarsky and Prof.~Vadim Cheianov for their kind hospitality at the Instituut Lorentz, Universiteit Leiden, The Netherlands, where the final part of this work was done.
\end{acknowledgments}

\end{document}